# Fundamental Understanding of Exposure and Process Chemistry for Enhanced Lithography and Stability of Metal Oxide Resists

Kevin M. Dorney[1], Ivan Pollentier[1], Fabian Holzmeier[1], Roberto Fallica[1], Ying-Lin Chen[1,2], Lorenzo Piatti[1,3], Dhirendra Singh[1], Laura Galleni[1,2], Michiel J. van Setten[1], Hyo Seon Suh[1], Danilo De Simone[1], Geoffrey Pourtois[1], Paul van der Heide[1], John Petersen[1]

[1]Imec, Kapeldreef 75, 3001 Leuven, Belgium
[2]Department of Chemistry, KU Leuven, Celestijnenlaan 200F, 3001 Leuven, Belgium
[3]Dipartinento di Elettronica e Telecomunicazioni, Politecnico de Torino, Corso Duca degli Abruzzi, 24, 10129 Torino, Italy

**ABSTRACT**

Metal oxide resists (MORs) have shown great promise for high resolution patterning in extreme ultraviolet (EUV) lithography, with potential for integration into high volume manufacturing. However, MORs have recently been shown to exhibit sensitivity to process conditions and environment, leading to critical dimension (CD) variation. While this variation can be reduced with proper process control, there is little knowledge on how these aspects affect the image formation mechanism. To bridge these knowledge gaps, we deploy a coordinated, fundamentals-focused approach to yield deep insights into MOR exposure and process chemistry. Our results on a model MOR, an n-butyl Sn-Oxo system, reveal how parameters such as exposure dose, post-exposure bake temperature, and atmospheric species influence the image formation mechanism. Our results, and the coordinated approach using correlative spectroscopies, provide a strong foundation for understanding the image formation mechanism in MOR materials with potential to link mechanistic aspects to CD variation.

## 1. INTRODUCTION

After decades of development, extreme ultraviolet (EUV) lithography is now revolutionizing high-volume manufacturing (HVM) of nanoelectronics by providing the possibility to print ever smaller features in today's and tomorrow's technology nodes. Recently, the ultimate scaling of lithography has been further reduced due to the advent of high-NA EUV lithography systems, which have recently shown an ultimate resolution of < 10 nm [1]. This extreme scaling of printing nanoscale features relies not only on advances in scanner hardware, optics (including EUV mask design/production), and computational lithography, but also on optimization of the imaging material itself, the photoresist. Metal oxide resist (MOR) platforms based on Sn-Ox systems have shown great promise for printing at the highest resolution possible, while also showing benefits of higher EUV absorption [2] and increased etch resist for pattern transfer [3]. The promise of MORs has resulted in an extensive investigation of their reactive pathways and lithographic performance over the last 10-15 years, which has resulted in not only a verification of their ability to serve as high-resolution resists, but also a generally accepted exposure mechanism [4-7]. When exposed to EUV light, these Sn-based MOR materials undergo ligand cleavage, resulting in the creation of "active sites" centered on the Sn atoms where ligand cleavage occurred. Following exposure, these activated sites can undergo a condensation reaction during the post-exposure bake (PEB) step, which results in a cross-linked, densified network that is retained during the development step (i.e., the MORs typically exhibit negative-tone behavior). This simplified reaction scheme, as compared to their chemical amplified resist counterparts, indicates that MORs have the potential for more straightforward modifications to improve their lithographic performance.

Recently, studies of the lithographic performance of MOR materials when accounting for delays in processing (e.g., post-coating and post-exposure delays, PCD and PED, respectively) have shown a variation in the critical dimension (CD) of the patterns observed after development [6-8]. These CD variations show sensitivity not only to the duration of the delay during processing [8], but also to the nature of the atmospheric environment during the delay step [6,7]. Such variations could be detrimental for process control in HVM applications and as such the use of MOR materials in HVM has, to date, been limited. While proper environmental control can reduce or even suppress delay-induced environmental reactions, the exact chemical nature of reactions of MOR materials with the environment remains elusive. Recently, it has been shown that the local environment during PEB can significantly influence the rate of ligand cleavage [9], which could influence the rate and degree of PEB-driven condensation due to either the creation of additional active sites or increasing the rate of thermolytic catalysis. Taken together, these findings show that environmental factors could add a degree of complication to the general description of chemical reactions underlying MOR lithographic performance. The fundamental understanding of these interactions presents a grand challenge for the lithographic community. On the one hand, a complete understanding would allow for informed design of MOR materials that suppress or eliminate this environmental sensitivity, thus increasing their process stability for HVM. On the other hand, such environmental interactions might be able to be leveraged to increase the performance of MORs, such as increasing ligand cleavage rates to create more active sites and thus (potentially) reduce the dose. To achieve these goals of increased stability and environmentally enhanced performance, fundamental investigations in process and exposure chemistry in MOR materials are required.

In this work, we utilize model MOR systems based on the $(SnR)_{12}O_{14}(OH)_6 \cdot X_2$ (R = ligands, X = counterion) platform [10] to perform investigations of the lithographic performance and environmental interactions that alter the processing chemistry. We leverage a combination of analytical techniques to investigate the exposure mechanism and show differences in dose and chemistry for model MOR materials with different counterions. To study effects of PED and PEB environmental conditions, we leverage correlative studies of FTIR and XPS to reveal how environmental interactions can alter the chemistry occurring in thin film MORs. Our results reveal the role of humidity and $O_2$ during PED and PEB, both on ligand cleavage and formation of Sn-O bonds. Furthermore, we investigate the environmental dependence of reactions occurring in *unexposed* regions of the MOR materials, which has consequences on the contrast and potential scumming after development. Our results not only shed light on these complex interactions but also provide foundational data to stimulate a fundamental understanding of environmental effects during MOR processing, which could eventually be leveraged for increased stability and improved performance.

## 2. EXPERIMENTAL

### 2.1 Sample preparation, characterization, and processing

Thin films of a model MOR material based on the familiar $(SnBu)_{12}O_{14}(OH)_6 \cdot X_2$ structure (Bu = butyl ligands, X = counterion) were prepared by spincoating the stock solutions (in either methyl iso-butyl carbinol or 1-butanol solvents) using instructions provided by the supplier. Depending on the experiment and material being used, the counterions could be composed of either $OH^-$ (hydroxide), $CH_3COO^-$ (acetate), or $(CH_3)_3COO^-$ (pivalate) ions (denoted by MOR-[OH], MOR-[Ac], and MOR-[Piv] in the following), and materials were either spincoated on 200mm wafers or 3 x 3 cm$^2$ coupons (see text for details). After spincoating, the films were subjected to a post-application bake (PAB) of 100 °C for 60 s to drive off any remaining solvent. The resulting films where then characterized via spectroscopic ellipsometry (RC2, JA Woollam company). The ellipsometry results indicated the spincoated films varied in thickness from ~21 – 26 nm (depending on the counterion and casting solvent) with a uniform surface roughness <0.5 nm. For EUV exposed samples where a PEB was applied, the PEB was performed for 60 s at a temperature of 200 °C in a standard cleanroom environment, unless otherwise noted. Samples that were investigated via Fourier transform infrared (FTIR) spectroscopy and X-ray photoelectron spectroscopy (XPS) were cleaved from double-side polished Si wafers into wafer pieces ('coupons') of ~ 3 x 3 cm$^2$, as the double side polishing of the substrate prevents scattering-induced signal loss during FTIR measurements. All samples were prepared on Si wafer substrates without application of a primer or underlayer.

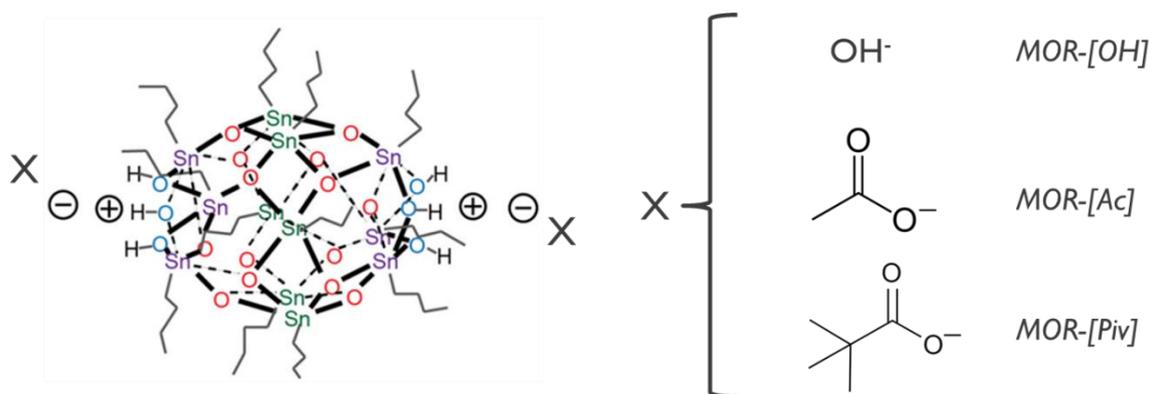

Figure 1. Model MORs used in this work, schematic of the Sn cage adapted from [11].

### 2.2 EUV Flood Exposures

EUV flood exposures of the model MOR materials were performed using two tools at imec; a spectrally filtered plasma-discharge-based z-pinch EUV source (EQ-10, Energetiq) integrated into imec's BEFORCE tool [9] and an EUV scanner (NXE3400B, ASML). For contrast curve measurements, flood exposures were performed over a 1x1 cm$^2$ field over a dose range of 0 to 170 mJ/cm$^2$ with a total of 77 exposures in each dose range. Dose increments were more closely spaced around the expected value of $E_0$ in order to better sample the contrast curve. Additional flood exposures were performed in imec's BEFORCE tool for samples that were investigated via residual gas analysis (RGA), FTIR, and XPS. To ensure a uniformly exposed area, the EUV spot (~5 mm in diameter) in the BEFORCE tool was raster scanned across an area of ~1.6 x 2.2 cm$^2$, at a speed that corresponded to a dose-to-gel (D2G) of 35 mJ/cm$^2$. Here, we define D2G as the dose obtained at 50% remaining resist thickness. The correspondence of the EUV dose between the flood exposure tool BEFORCE and the NXE scanner has been calibrated and compared using the same model MOR materials.

### 2.3 EUV-Induced Desorption Mass Spectrometry for Residual Gas Analysis

RGA measurements are performed in the operational mode of EUV-induced desorption mass spectrometry, whereby molecular fragments released from the model MOR materials during EUV exposure are ionized by a filament and then mass filtered before being detected by an ion detector. The measurement of outgassed species via imec's RGA tool has been described previously [12], but the process is briefly summarized here. The model MOR materials were spincoated onto 200 mm wafers using recipes provided by the supplier. A PAB of 100° C was applied for 60 s to drive off any residual solvent and the wafers were loaded into the BEFORCE tool for RGA analysis. EUV exposures were performed in vacuum ($10^{-6}$ mbar) and the spot was raster scanned over the wafer with a dwell time that corresponded to an exposure dose of ~7 mJ/cm$^2$. Molecular fragments were collected with a Pfeiffer QMG422 quadrupole mass spectrometer over the range of 1 – 150 amu, with a resolution of 1 amu.

### 2.4 Fourier Transform Infrared Spectroscopy in a Controlled Environment

Measurements of the IR absorption spectrum of the model MOR materials were performed using an iG50 FTIR spectrometer (ThermoFisher) in transmission mode, over a range of 600 – 4,000 cm$^{-1}$ with a resolution of 4 cm$^{-1}$. A baseline IR absorption spectrum was recorded on each sample to establish the unexposed chemical nature of the model MORs. An uncoated double-polished Si wafer from the same lot was used for the background spectrum. Following the initial measurement, the samples were loaded into the exposure chamber of the BEFORCE tool and after EUV exposure they were transferred under vacuum (~$10^{-3}$ mbar) and then stored (also in vacuum) until further measurement and processing. Measurements of the post-EUV exposure IR absorption spectrum were performed by transferring the exposed samples to a custom-built goniometer (also under vacuum), which allows for setting an arbitrary angle of incidence of the IR beam with respect to the sample plane. Following post-exposure measurements, the samples were then transferred to the BEFORCE PEB chamber, where PEBs were carried out under custom-defined atmospheres (see Results section for more details). After PEB, a final IR spectrum was recorded by transferring the baked sample back to the goniometer. For all measurements in this study, an angle of 30 degrees of the IR beam with respect to the sample plane was chosen to increase the effective path length of the model MOR materials. Furthermore, the vacuum level in the FTIR sample chamber was <$10^{-1}$ mbar for all measurements.

**2.5 X-ray Photoelectron Spectroscopy**

X-ray photoelectron spectroscopy (XPS) measurements were performed using a QUANTES XPS tool (Physical Electronics, PHI) employing a monochromatized Al-Kα source (1486.6 eV) with a spot size of ∼100 μm. The spot was then raster scanned over an area of 500 x 1000 μm$^2$ to reduce the X-ray dose and prevent further chemical changes during measurement [13]. Samples of the exposed and baked model MOR materials were cleaved to $1 \times 1$ cm$^2$ coupons and XPS measurements were made at the C 1s, O 1s, and Sn 3d absorption edges (which covers all atomic species in the MOR). An electron flood gun was employed for charge neutralization during the measurements to prevent charging-induced peak shifts from the insulating nature of the MOR material.

Peak deconvolution of the C 1s and O 1s regions of the XPS spectrum were performed using the lmfit package as implemented in Python [14]. Prior to peak deconvolution, a Tougaard background [15] was applied and subtracted from the raw spectra to remove contributions from photoelectron scattering. For the resulting peak deconvolution, a mixed set of Gaussian-Lorentzian peaks (90/10) were used to account for effects of different broadening mechanisms in the XPS spectra [16] as well as finite lifetime of the core-hole states. For all fitting, the peak center positions were fixed to within 0.1% of the experimentally measured peak positions, while the widths and amplitudes were adjusted to obtain a best fit of the data. In all cases, an R-squared value >0.99 is obtained from the fits.

## 3. RESULTS AND DISCUSSION

*3.1 Residual outgas analysis of EUV-induced ligand cleavage in model MOR films*

Upon exposure to EUV light the model Sn-Ox model cage system undergoes ligand cleavage, which can be further enhanced via interaction with primary and secondary electrons [17] or thermal treatment (e.g., a PEB) [5,6]. The exact mechanism of the ligand cleavage process has been debated in literature, with recent simulations on methyl-substituted clusters suggesting homolytic cleavage as the energetically favored pathway for ligand removal [18]. Homolytic cleavage would result in a radical cation Sn cluster —centered on the Sn site where ligand cleavage occurred— and an alkyl radical; however, radicals are difficult to detect in the film and thus confirmation of the homolytic cleavage process remains elusive. To investigate the ligand cleavage process, we performed RGA measurements on the MOR-[Ac] model material, and the recorded mass spectrum resulting from EUV-induced desorption is shown in Figure 2.

Since EUV exposure does not result in significant removal of Sn or O species, the mass spectrum is predominately composed of products arising from the butyl ligand. Of particular interest in the mass spectrum are two signals at m/z = 56 and 114 (pink bars, Figure 2a), which correspond to 1-butene and octane, respectively. The presence of octane in the mass spectrum is a strong indication of butyl radical recombination, which (as a necessity) would require homolytic cleavage of the Sn-C bonds in the cluster to produce the butyl radicals. Furthermore, the signals and m/z values corresponding to fragmentation products of the octane molecule are also observed (red bars, Figure 2a and 2b) and give further support of the presence of octane in the outgassed species during EUV exposure. Interestingly, the strong signal at m/z 56 indicates hydrogen abstraction from the cleaved butyl ligand, which itself could be donated to the radical Sn site and form a tin hydride bond, as suggested previously for deep UV (DUV) irradiation [11]. Such a mechanism would also suggest a pseudo β-hydride elimination reaction, which would be enhanced via a greater density of β-hydrogens in the ligand (e.g., an isopropyl ligand group). However, evidence of the Sn-H bond formation has not, to our knowledge, been conclusively confirmed. Nonetheless, the RGA measurements provide strong evidence that ligand cleavage occurs via homolytic radiolysis and results a in radical cation centered on an Sn atom, which could form the basis of the "active site" of these compounds.

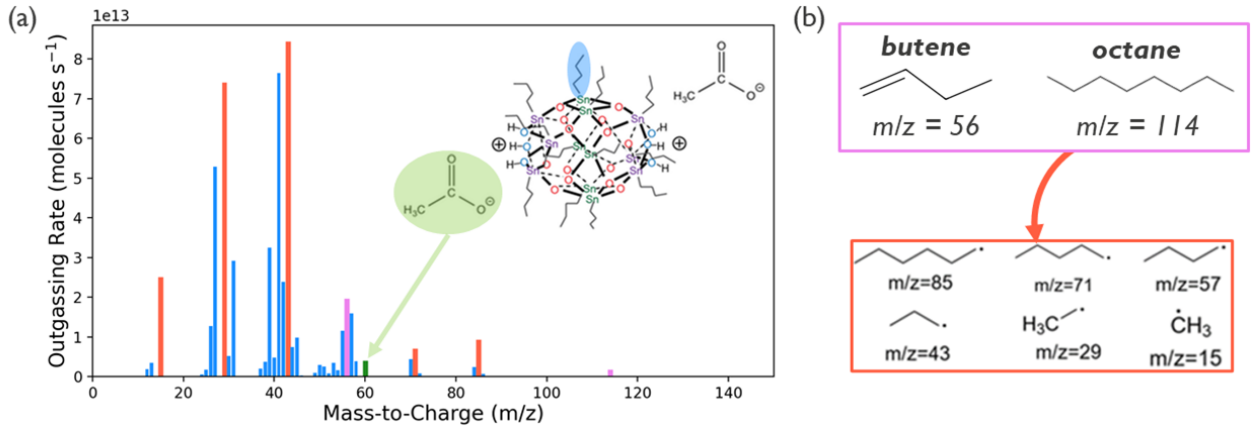

Figure 2. Mass spectrum recorded during EUV exposure of the MOR-[Ac] material together with peak assignment. (a) Converted mass spectrum into units of outgassing rate (molecules sec$^{-1}$) as a function of detected mass. Most of the signal results from ionization and fragmentation of the butyl ligands, highlighted in blue in the schematic. A small signal at m/z=60 is observed, which is assigned to the acidic form of the [Ac] counterion (green). (b) Structure of the butene and octane molecules, which correspond to the signals detected at m/z=56 and 114 in panel (a). In red, the fragmentation pathways of the octane molecule are shown, which match well to the signals observed in the mass spectrum. The inset in panel (a) shows a schematic of the MOR-[Ac] material

## 3.2 Effects of counterions on lithographic contrast in model MOR

Having established the role of homolytic cleavage resulting from EUV exposure, we next investigated the role of the counterion in a similar fashion via RGA and FTIR analysis. While initial investigations show little dependence of counterion properties (e.g., bond dissociation energy or molecular mass) [10] on lithographic performance, recent studies have shown that the $E_0$ values of the SnBu clusters can be significantly enhanced or suppressed via the choice of the counterion [19]. To investigate this effect in a controlled manner, we measured the contrast curves of three butyl-terminated model MOR clusters with hydroxide (OH$^-$), acetate (CH$_3$COO$^-$), and pivalate ((CH$_3$)$_3$COO$^-$) counterions, which are shown in Figure 3. As can be seen from the measured contrast curves on imec's NXE 3400B scanner, there is a significant dose shift between the three different counterions. The exposure parameters of D2G and lithographic contrast, γ, are determined by fits of the contrast curve data to the following equation,

$$y = a_2 + \frac{a_1 - a_2}{1 + \left(\frac{x}{x_0}\right)^\gamma}$$  [1]

where $y$ is the resist thickness after development, $x$ is the exposure dose, $x_0$ is the D2G (at 50% remaining resist thickness), γ is proportional to lithographic contrast, and $a_1$ and $a_2$ correspond to the resist thickness at 0 and infinite dose, respectively.

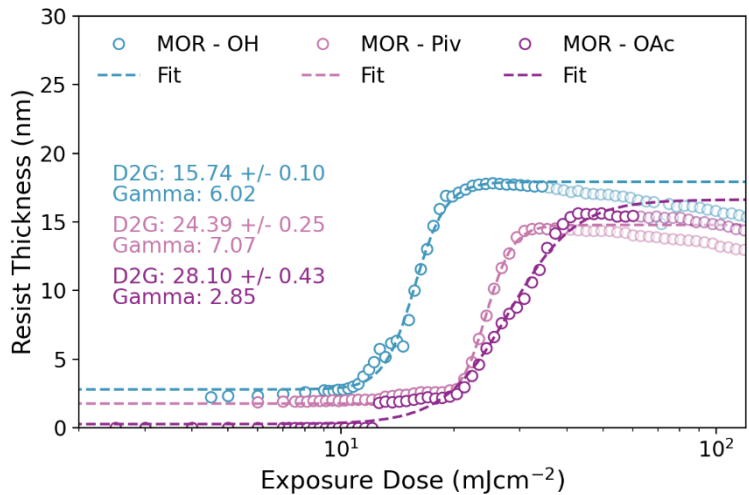

Figure 3. EUV contrast curves measured on the three model MOR materials with no underlayer. The results of the fitting of the contrast curves via Eq [1] are shown as an inset in the plot, where D2G is the dose-to-gel at 50% remaining thickness. Resist thicknesses values obtained at a high dose where resist "shrinkage" was observed are not included in the fitting (light colored markers).

To further investigate the effects of the counterions on the EUV exposure dynamics, RGA measurements were performed on the MOR-[Ac] and MOR-[Piv] materials. Aside from the typical mass fragments that are observed for the butyl-substituted cluster (cf. Figure 2), signals at m/z 60 (MOR-[Ac]) and 102 (MOR-[Piv]) are detected, as well as an enhanced signal at m/z 57, which corresponds to the tert-butyl group of the pivalate counterion. These mass fragments correspond to the acidic form of the two counterions, suggesting that the counterions themselves may extract protons from the Sn cluster (or perhaps the butyl radicals themselves). Interestingly, it has been shown that strong acids (e.g., HBr) can react with Sn clusters and result in ligand cleavage [20], which could also indicate that weaker organic acids may also interact with the Sn cluster during and following exposure. However, we do note that FTIR measurements have indicated a relatively similar amount of CH stretching signal loss for the three different counterions (Figure 4), which suggests that differences in D2G are due to post-exposure interactions. However, further investigations are needed to elucidate the chemical interaction of counterions during and after EUV exposure in these model MOR systems.

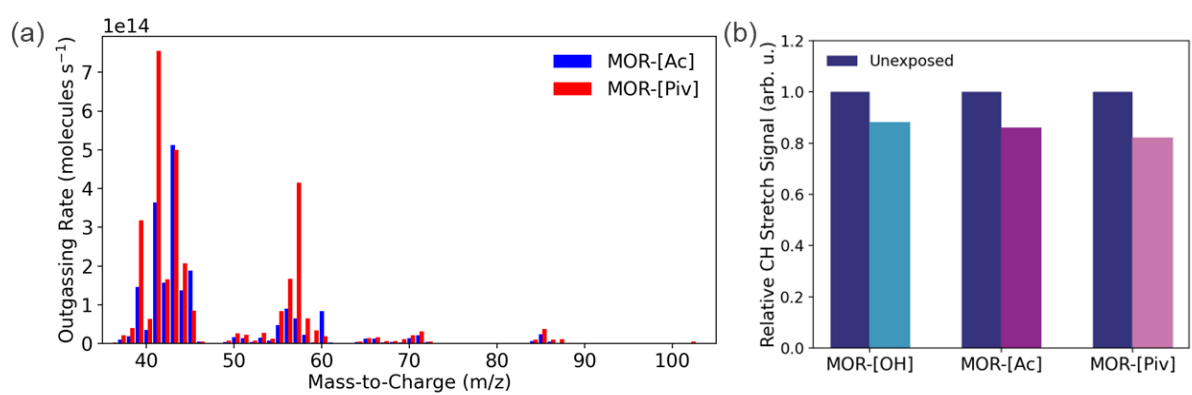

Figure 4. (a) Mass spectrum recorded during exposure of the MOR-[Ac] and MOR-[Piv] films. While many of the peaks correspond to signals from the butyl ligands, signals at m/z 60 and m/z 102 are related to the acidic forms of the counterions (acetic acid and pivalic acid, respectively). Additionally, an enhanced signal at m/z 57 is observed for pivalate, that could be explained by cleavage of the tert-butyl group. (b) Relative ligand signal remaining in the model MOR films before (dark blue) and after an exposure of 35 mJ/cm$^2$ (light colors, same as in Figure 3). The change in ligand signal is relatively similar for the 3 films, suggesting these counterions do not significantly affect the ligand cleavage rate.

*3.3 Interactions of model MOR with the environment during post-exposure delay (PED)*

It has been recently shown that commercial MOR materials can rapidly uptake $H_2O$ during post-exposure delays, which can affect the lithographic performance [6,7]. However, the exact chemical interaction between exposed MOR materials and the atmospheric environment remains an open question. To better understand the chemical interactions following post-exposure delay, a first experiment was performed where the MOR-[Ac] material was exposed at varying exposure doses of 35, 50, and 100 mJ/cm$^2$ and then subjected to XPS analysis to reveal chemical changes occurring within the film upon exposure to standard cleanroom atmosphere. Exemplary XPS spectra recorded at the C 1s and O 1s absorption edges are shown in Figure 5a and 5b, along with the corresponding sub-peaks obtained via the deconvolution routine. The trends in sub-peak area through exposure dose, which represent a change in chemical content, are shown in Figures 5c and 5d.

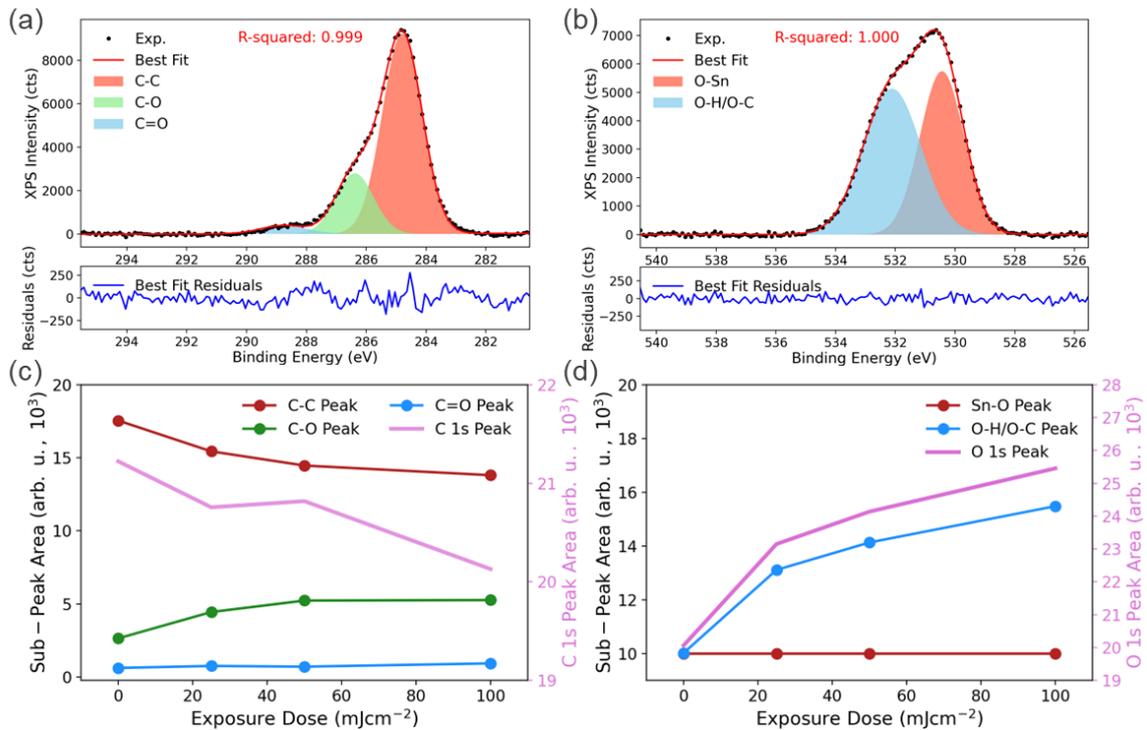

Figure 5. XPS analysis of the MOR-[Ac] material that was subjected to a long PED at varying exposure dose. Raw XPS spectra collected at the C 1s (a) and O 1s (b) absorption edges, where the underlying sub-peaks (shaded regions) contributing the experimental signal (black dots) signal are shown. The best fit from the peak deconvolution routine is shown as the red line, while residuals (defined as $I_{exp} - I_{fit}$) are plotted below the main signal. Sub-peak areas along with total integrated peak signal as a function of exposure dose for the (c) C 1s and (d) O 1s peak.

For the C 1s spectra, we see an overall decrease of the sub-peak corresponding to C-C bonds, which is a result of exposure-induced ligand cleavage and the main contributor to the total C 1s signal as indicated by the decrease in total C 1s peak area (pink line, Figure 5c). However, we also observe an increase in the C-O sub-peak with respect to exposure dose, which suggests that interaction of exposed Sn clusters with atmosphere results in the formation of oxygenated carbon species. The C=O sub-peak shows only a weak dependence on exposure dose; however, this apparently weak dependence is likely convoluted by the loss of acetate counterions with increasing exposure dose and the formation of oxygenated carbon products as indicated by the increase in C-O peak.

For the O 1s peak, we observe a strong increase in the O-H/O-C sub-peak area (these subpeaks cannot be reliably separated, so are treated as one sub-peak) as a function of exposure dose, while the O-Sn sub-peak at lower binding energy remains relatively constant. This strong increase is likely the result of the combined effect of $H_2O$ accumulation in the film and the formation of oxygenated carbon by-products. Taken together, the XPS spectra indicate that activated Sn sites resulting from EUV exposure can interact with atmospheric gases and $H_2O$, which modify the film chemistry by the formation of new chemical bonds.

*3.4 The role of humidity and oxygen during PEB on MOR chemistry*

Having identified the uptake of $H_2O$ during PED and potential chemical modifications of the exposed MOR film during PED and PEB, we next investigated the role of $H_2O$ and $O_2$ during the PEB step, as $O_2$ is the next largest component in atmospheric gases (aside from neutral $N_2$). While pure $O_2$ gas streams are not yet available in our BEFORCE tool, we isolate the role of $O_2$ via a comparative measurement using $N_2$ and clean dry air (CA) flow streams at the same relative humidities. In this configuration, the main differences between the two gas streams are the approximately 21% of $O_2$ and ~0.05% of $CO_2$ in the CA stream. Owing to the relatively minor amount of $CO_2$ in the CA stream, we attribute the main differences in the resulting chemical changes during PEB to be coming from the presence of $O_2$. Figure 6a shows the relative CH stretching signal (calculated as the integrated area of the CH stretching peak from 2,700 – 3,100 cm$^{-1}$) of the MOR-[Ac] material as a function of relative humidity (RH) between the two gas streams (the humidity is controlled via humidifier between the gas input and the PEB chamber). The samples are exposed with the same EUV exposure dose of 35 mJ/cm$^2$ and subjected to the same PEB temperature and time (200 °C for 60 s) and with a PED of ~15 s in N$_2$.

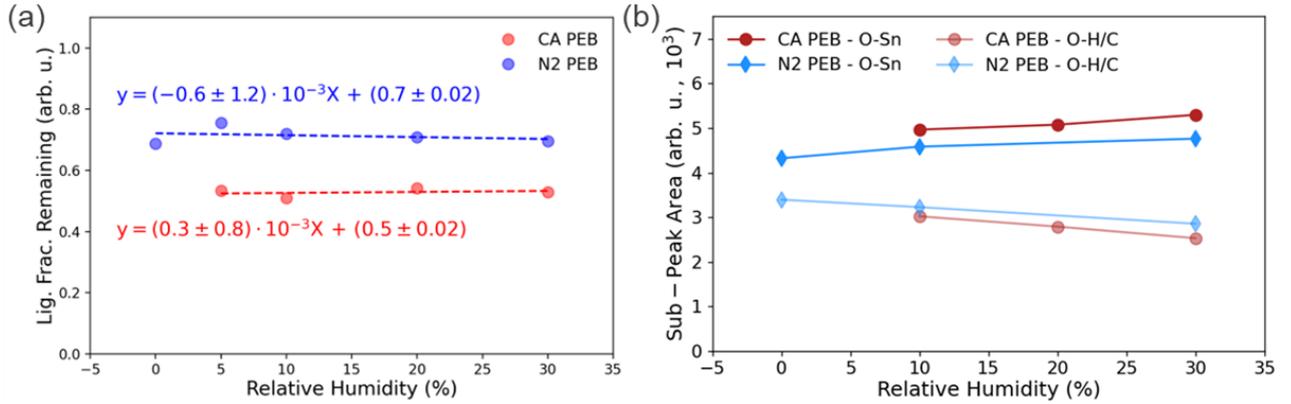

Figure 6. MOR chemistry (ligand cleavage rates and O-Sn formation/O-H/C loss) as a function of relative humidity (RH) during PEB (200 °C, 60 s) for CA and N$_2$ environments at an exposure dose of 35 mJ/cm$^2$. (a) Relative ligand fraction remaining in the MOR film as function of RH. Linear fits to the data show no dependence on the RH for the ligand fraction cleaved during PEB. (b) Integrated areas of O-Sn and O-H/C peaks as a function of RH for the two PEB environments. A slight, linear dependence on RH is observed for both environments, while an increased rate of Sn-O formation is observed for CA compared to N$_2$ environment.

From the relative ligand signal remaining in the film, there are two clear observations. Firstly, the ligand cleavage rate does not exhibit a dependence on the RH during PEB over the range of RHs studied (as evidenced by the poor linear fit and large error in slope), regardless of the two gas streams. Secondly, an enhanced rate of ligand cleavage is observed for the PEB performed in the CA environment across all RH values, resulting in ~40-50% more ligand loss as compared to the N$_2$ PEB environment. This large difference is likely the result of the presence of O$_2$, which participates during the thermocatalytic reaction and leads to a larger amount of ligand cleavage. While studies of the effect of O$_2$ during exposure and PEB are currently lacking, we do note a study on a Keggin-type NaSn cluster showed enhanced butyl ligand loss during e-beam exposure, even at very low O$_2$ partial pressures of 10$^{-7}$ Torr [21].

While O$_2$ can enhance the ligand cleavage rate, the contrast mechanism in MOR materials also requires the formation of Sn-O-Sn linkages so that an insoluble, dense network can be formed for transfer the EUV pattern into the underlying substrate layer. The formation of Sn-O-Sn linkages is believed to form via condensation reactions during PEB and thus should require H$_2$O and also exhibit a dependence on the RH of the PEB environment. To elucidate this dependence, we subjected the same post-PEB samples to XPS analysis to quantify the formation of Sn-O-Sn as well as the expected loss of O-H and O-C bonds. The integrated areas of the sub-peaks at the O 1s edge (Figure 6b) indeed show an increase in Sn-O-Sn content as a function of increasing RH for both gas streams, while also showing an increased offset for CA gas stream. These trends are also reflected in the loss of O-H/O-C as a function of increasing RH, which further supports the formation of Sn-O-Sn linkages via a thermocatalytic process that is enhanced in an O$_2$ environment.

### 3.5 Effects of PEB environment on the contrast mechanism in model MOR

In the previous sections, we thoroughly investigated the interaction of atmospheric species with an exposed model MOR material; however, the variation in litho performance as observed with proprietary MOR materials is, from a fundamental viewpoint, due to a variation in the latent chemical image. Since the chemical image depends upon strict control of exposed vs unexposed regions, a complete understanding of the environmental interactions and their effect on contrast demands a study of interactions in unexposed regions as well. Moreover, since difference environment can lead to different changes in ligand cleavage rates [9], it is possible that the dose-to-gel also shifts depending on PEB environment. As such, studies of ligand cleavage and SnO formation should be performed through exposure dose and through PEB temperature to better understand the effects of environment on the contrast mechanism.

To study these effects, we interrogated both exposed and unexposed model MOR films in two ways; for exposed MOR materials, a series of exposure doses were performed and then followed by a PEB (200 °C, 60s) while for unexposed MOR materials, a series of bake temperatures were applied (to mimic chemical changes that would occur during a PEB of a patterned sample). This "thermolysis bake" follows simple first-order kinetics with respect to the ligand removal rate as a function of temperature [5], thus enabling a quantitative mechanistic comparison between the different environments. These experiments were then performed in CA and $N_2$ flow streams with 30% RH.

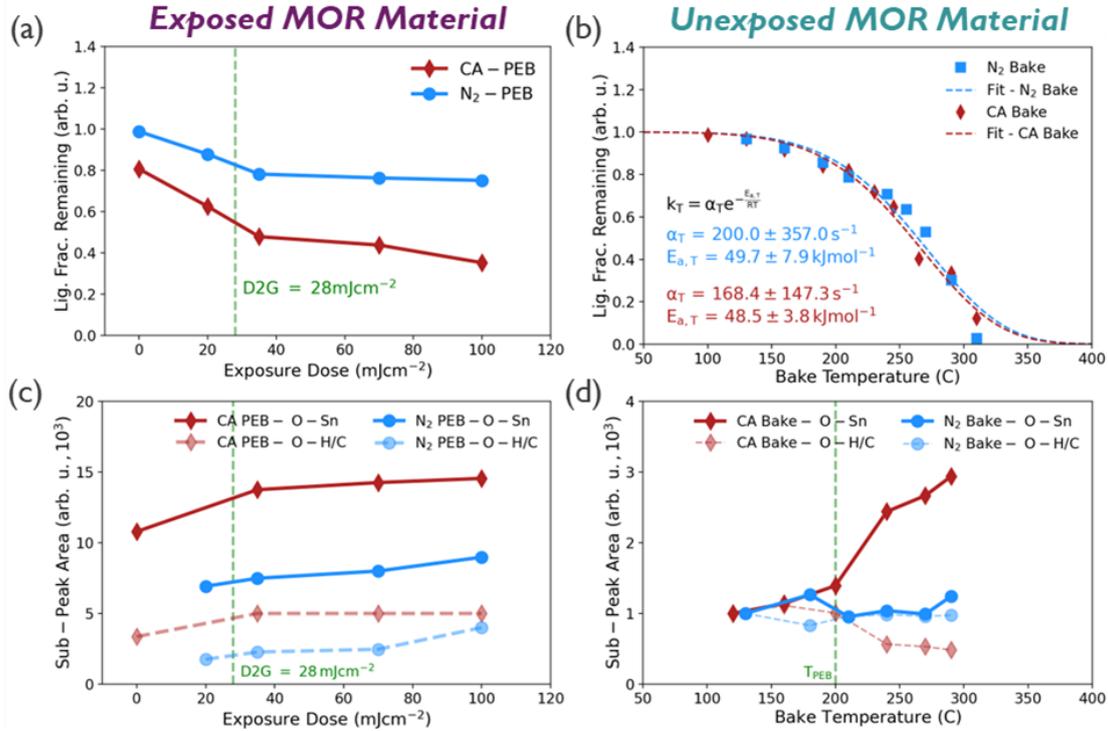

Figure 7. Comparison of the ligand fraction remaining and the SnO formation in exposed (a,c) and unexposed (b,d) MOR-[Ac] for PEB environments in clean air (CA) and nitrogen ($N_2$). For the exposed MOR, the ligand fraction remaining is significantly reduced for the CA PEB, while the formation of Sn-O bonds is also enhanced as indicated by XPS measurements (c). For the unexposed region, the remaining ligand fraction is similar over the range of temperatures studied for both CA and $N_2$. However, SnO formation still occurs in unexposed regions even without exposure in CA. For bakes performed in $N_2$ the formation of SnO is significantly suppressed, even at high temperatures.

In Figure 7, we show the combined FTIR (7a and 7b) and XPS (7c and 7d) results for the exposed and unexposed MOR-[Ac] material in CA and $N_2$ environments. For the exposed MOR-[Ac], we observe a nearly 50% increase in the amount of ligands removed across all exposure doses, with the greatest change in ligand removal being observed just before the nominal D2G (as determined from contrast curve measurements on the same exposure tool). Interesting, after the D2G, the removal of additional ligands plateaus, with only a minor increase with respect to exposure dose. Nonetheless, the two curves for the ligand fraction remaining follow a similar trend for the two PEB environments, suggesting that the environment merely causes a near uniform offset of the ligand cleavage rates with respect to exposure dose. For the SnO formation, the XPS results show an increased amount of SnO formation as compared to the $N_2$ environment, also occurring across all exposure doses. Taken together, these results indicate that a CA environment results in both a higher amount of ligands being cleaved as well as Sn-O-Sn linkages being formed during the PEB. Given the main difference between CA and $N_2$ is the ~21% $O_2$ in the CA stream, these results also strongly suggest that chemical reactions in exposed regions of MOR can be enhanced via $O_2$.

In terms of the unexposed MOR material, we find the kinetic rate (in both activation energy, $E_{\alpha,T}$, and Arrhenius pre-factor, $\alpha$) for ligand removal is independent of whether the thermolysis bake is performed in CA or N environments. This suggests that ligand cleavage via the thermolysis reaction is neither enhanced nor suppressed based on the PEB environment. As such, it may be possible to enhance ligand cleavage in exposed regions via PEB environments, while not altering the ligand cleavage rates in unexposed regions. However, the XPS results show that at elevated temperatures, Sn-O-Sn linkages can be formed in unexposed regions for the CA PEB, which would potentially manifest as scumming or other defects in a patterned sample. While more optimization and further testing is needed to fully explore the effects of environment on exposed and unexposed MOR, these results present a promising route to leveraging the PEB environment to selectively tune ligand cleavage and SnO formation in exposed/unexposed regions.

## 4. CONCLUSION

In this work, we have utilized model MOR materials based on the $(SnBu)_{12}O_{14}(OH)_6 \cdot X_2$ structure, together with advanced characterization techniques, to reveal fundamental mechanistic aspects of exposure and process chemistry. The initial reaction in the exposure mechanism, the cleavage of ligands from the Sn cage, has been shown with a high probability to occur via homolytic cleavage and lead to the production of butyl radicals. Additionally, EUV exposure liberates counterions from the film and, in the case of acetate and pivalate, lead to further interactions (e.g, protonation). While the interaction of the counterion with the resist remains an area to be explored, our results suggest that the counterion may need to be considered as more than just a passive charge neutralization species and rather as a component that influences MOR chemistry.

In addition to fundamental exposure process, we have also extensively investigated the subsequent process chemistry occurring during PED and PEB. During PED, water can quickly interact with the exposed MOR material, but in addition the formation of oxygenated carbon by-products can also occur. The exact chemical nature of these interactions remains to be investigated further but is indicative that non-reversible interactions could influence MOR chemistry during PED. Furthermore, we extensively studied the effect of PEB environment on MOR chemistry, showing that a) ligand cleavage is relatively insensitive to relative humidity during PEB, b) $O_2$ appears to enhance both ligand cleavage and enable higher rates of Sn-O-Sn formation, c) ligand cleavage in unexposed MOR is relatively independent of PEB environment, and d) that Sn-O-Sn formation in unexposed regions can be suppressed with $N_2$ during PEB. While these results represent only a subset of the complex environmentally mediated reaction occurring during MOR exposure and processing, they provide a foundation for understanding fundamental chemical modifications and potentially a route for leveraging these reactions for increased performance and stability of MORs. Finally, we note that while commercial MOR materials might exhibit qualitative differences in response to environment, exposure dose, and PEB conditions as compared to the model MORs studied here, we believe a correlative approach using advanced fundamental characterization techniques provides a powerful platform for unraveling their complex chemistries.

## 5. ACKNOWLEDGEMENTS

The authors like to thank Intel and resist suppliers for providing model MOR materials. Special thanks also to James Blackwell, Eric Mattson, and Charles Mokhtarzadeh (Intel) for support and helpful discussion. This work has been enabled in part by the NanoIC pilot line. The acquisition and operation are jointly funded by the Chips Joint Undertaking, through the European Union's Digital Europe (101183266) and Horizon Europe programs (101183277), as well as by the participating states Belgium (Flanders), France, Germany, Finland, Ireland and Romania. For more information, visit www.nanoic-project.eu.

## REFERENCES

[1] Blanco, V., et al., "Logic and memory patterning breakthrough in the imec ASML High-NA lab," Proc. SPIE, 13216, 119 (2024). https://doi.org/10.1117/12.3047176

[2] Fallica, R., et al., "Absorption coefficient of metal-containing photoresists in the extreme ultraviolet," J. Micro/Nanopatterning Mater. Metrol., 17 (2), 023505 (2018). https://doi.org/10.1117/1.JMM.17.2.023505

[3] Stowers, J., et al., "Metal oxide EUV photoresist performance for N7 relevant patterns and processes", Proc. SPIE 9779, Advances in Patterning Materials and Processes XXXIII, 977904 (2016). https://doi.org/10.1117/12.2219527

[4] De Schepper, P., et al., "MOx resist formulation and process advances towards high-NA EUV lithography,", Proc. SPIE 12498, Advances in Patterning Materials and Processes XL, (2023). https://doi.org/10.1117/12.2658499

[5] Needham, C., et al., "Advanced simulations using an improved metal oxide photoresist model", Proc. SPIE 12957, Advances in Patterning Materials and Processes XLI, 129571B (2024). https://doi.org/10.1117/12.3010941

[6] Castellanos, S., et al., "EUV metal oxide resists: impact of the environment composition on CD during post-exposure delay", Proc. SPIE 12957, Advances in Patterning Materials and Processes XLI, 1295707 (2024). https://doi.org/10.1117/12.3010921

[7] San Roman, R. R. M., et al. "Stable stitching: impact and mitigation of environment on metal-oxide resist imaging", J. Micro/Nanopatterning Mater. Metrol., 24 (1), 011011 (2025). https://doi.org/10.1117/1.JMM.24.1.011011

[8] Kim, S., et al., "An investigation on the process control for the solid application of EUV MOR", Proc. SPIE 12494, Optical and EUV Nanolithography XXXVI, 124940V (2023). https://doi.org/10.1117/12.2658345

[9] Pollentier, I., Holzmeier, F., Dorney, K., Suh, H. S., "BEFORCE: a pathway to unravel metal oxide resist (MOR) reactions upon EUV exposure, bake, and environment" Proc. SPIE 13215, International Conference on Extreme Ultraviolet Lithography 2024, 132150G (2024). https://doi.org/10.1117/12.3034708

[10] Cardineau, B., "Photolithographic properties of tin-oxo clusters using extreme ultraviolet light (13.5 nm)" Microelectron. Eng. 127 (5), 44 (2014). https://doi.org/10.1016/j.mee.2014.04.024

[11] Zhang, et al., "Photochemical conversion of tin-oxo cage compounds studied using hard x-ray photoelectron spectroscopy", J. Micro/Nanopatterning Mater. Metrol., 16 (2), 023510 (2017). https://doi.org/10.1117/1.JMM.16.2.023510

[12] Pollentier, I., et al. "Unraveling the role of photons and electrons upon their chemical interaction with photoresist during EUV exposure", Proc. SPIE 10586, Advances in Patterning Materials and Processes XXXV, 105860C (2018). https://doi.org/10.1117/12.2299593

[13] Sajjadian, F. S., et al., "Photoemission spectroscopy on photoresist materials: A protocol for analysis of radiation sensitive materials", J. Vac. Sci. Technol. A 41, 053206 (2023). https://doi.org/10.1116/6.0002808

[14] Newville, M. et al., (2024) lmfit/lmfit-py: 1.3.2 (1.3.2). Zenodo. https://doi.org/10.5281/zenodo.12785036

[15] Tougaard, S. "Practical guide to the use of backgrounds in quantitative XPS" J. Vac. Sci. Technol. A 39, 011201 (2021). https://doi.org/10.1116/6.0000661

[16] Galleni, et al. "Peak Broadening in Photoelectron Spectroscopy of Amorphous Polymers: The Leading Role of the Electrostatic Landscape" J. Phys. Chem. Lett. 15 (3), 834-839 (2024). https://doi.org/10.1021/acs.jpclett.3c02640

[17] Bespalov, I., et al., "Key Role of Very Low Energy Electrons in Tin-Based Molecular Resists for Extreme Ultraviolet Nanolithography", ACS Appl. Mater. Interfaces 12 (8), 9881 – 9889 (2020). https://doi.org/10.1021/acsami.9b19004

[18] Ma, J. H., et al., "Mechanistic Advantages of Organotin Molecular EUV Photoresists", ACS Appl. Mater. Interfaces 14 (4), 5514 – 5524 (2022). https://doi.org/10.1021/acsami.1c12411

[19] Kang, Y. K., et al., "Enhancement of photosensitivity and stability of Sn-12 EUV resist by integrating photoactive nitrate anion", Appl. Surf. Sci. 656 (30), 159564 (2024). https://doi.org/10.1016/j.apsusc.2024.159564

[20] Kenane, N., et al., "Dry Deposition and Dry Development of Metal Oxide Based Photoresist", J. Photopolym. Sci. Tec. 37 (3), 257 – 262 (2024). https://doi.org/10.2494/photopolymer.37.257

[21] Frederick, R. T., et al., "Effect of Oxygen on Thermal and Radiation-Induced Chemistries in a Model Organotin Photoresist", ACS Appl. Mater. Interfaces 11 (4), 4514 – 4522 (2019). https://doi.org/10.1021/acsami.8b16048